\documentclass[twocolumn,showpacs,superscriptaddress]{revtex4}
\usepackage{amsmath,graphicx,epsfig}

\begin{document}

\title{Polarized alkali vapor with minute-long transverse spin-relaxation time} 

\author{M. V. Balabas}
\address{S. I. Vavilov State Optical Institute, St. Petersburg, 199034 Russia}
\author{T. Karaulanov}
\address{Department of Physics, University of California at
Berkeley, Berkeley, California 94720-7300}
\author{M.\ P.\ Ledbetter}\email{ledbetter@berkeley.edu}
\address{Department of Physics, University of California at
Berkeley, Berkeley, California 94720-7300}
\author{D.\ Budker}
\address{Department of Physics, University of California at
Berkeley, Berkeley, California 94720-7300}
\address{Nuclear Science Division, Lawrence Berkeley National
Laboratory, Berkeley CA 94720}

\date{\today}


\begin{abstract}
We demonstrate lifetimes of atomic populations and coherences in excess of 60 seconds in alkali vapor cells with inner walls coated with an alkene material.  This represents two orders of magnitude improvement over the best paraffin coatings. Such anti-relaxation properties will likely lead to substantial
improvements in atomic clocks, magnetometers, quantum memory, and
enable sensitive studies of collisional effects and
precision measurements of fundamental symmetries.
\end{abstract}




\maketitle

Long-lived ground-state coherences in atomic vapor cells form the
basis for atomic clocks \cite{Rob1982,Ris1980,Rah1987},
magnetometers \cite{Ale1992,Bud2007}, quantum memory \cite{Jul2004},
spin-squeezing and quantum non-demolition measurements
\cite{Kuz2000,Was2010}, and precision measurements of fundamental
symmetries \cite{Gri2009}. One method for achieving long coherence
times is to coat the walls of a cell with an anti-relaxation film
such as paraffin \cite{Rob1958,Bou1966} or octadecyltrichlorosilane
\cite{Sel2007}.  Conventional paraffin coatings are formed from
long-chain alkane molecules, supporting approximately $10^4$
atom-wall collisions before depolarizing the alkali spins. In this
Letter we report on the remarkable anti-relaxation properties of a
new, alkene based, coating. With proper experimental arrangements,
we realize coherence lifetimes on the order of 1 minute in a 3 cm
diameter cell, corresponding to about $10^6$ polarization preserving
bounces.  To the best of our knowledge, this corresponds to the
narrowest electron paramagnetic resonance ever observed.

One of the key ingredients to realizing such long lifetimes is to
work in magnetic fields such that the Larmor precession frequency is
small compared to the spin-exchange rate, and to optically pump the
alkali vapor with circularly polarized light.  This largely
eliminates relaxation due to spin-exchange collisions, the so called
spin-exchange relaxation-free (SERF) regime \cite{All02,Hap1973}.
SERF magnetometers presently hold the record for magnetic field
sensitivity of any device \cite{Kominis2003,Kor2007}, but usually
require operation at temperatures in excess of $150{\rm ^\circ C}$.
The alkene coating described here enables operation of such a
magnetometer in a room temperature environment, dramatically
expanding its useful range of application, especially where low
power consumption is important. We present an experimental and
theoretical investigation of a room temperature atomic magnetometer
operating in the SERF regime. Experiment and theory are in good
agreement with each other.

\begin{figure}
  \includegraphics[width=3.4 in]{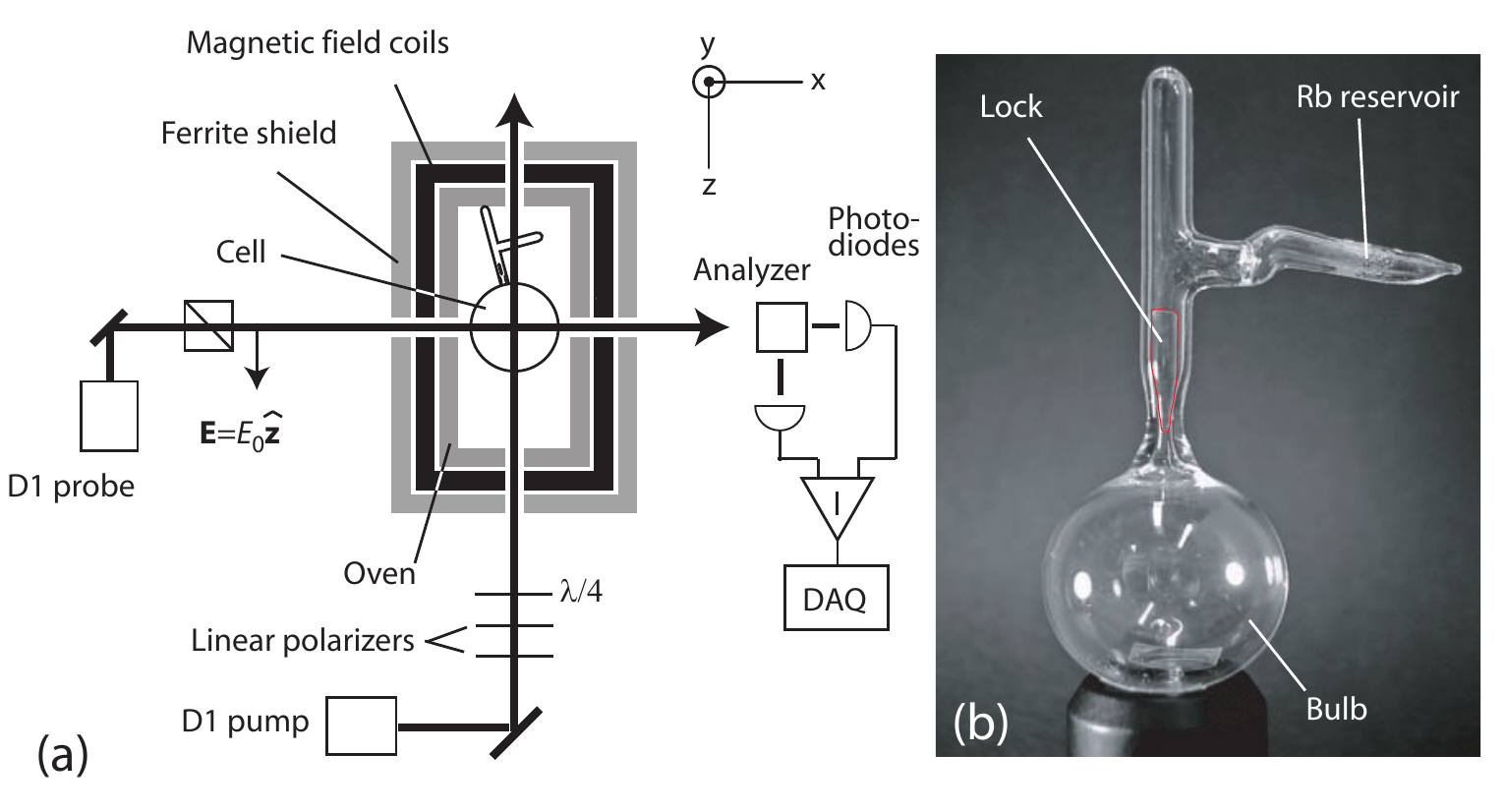}\\
  \caption{Experimental setup:  (a) Four mu-metal shields (not shown) surround the ferrite shield.
  The ferrite shield is a cylinder, 15.2 cm long and 11.4 cm in diameter (inner dimensions) with 1.3
  cm thick walls. A set of coils provides control over the magnetic fields.
  An oven, heated by a twisted pair of 30 gauge copper wire and AC current, provided temperature
control. (b) Photo of the coated cell.
 }\label{Fig:exp_setup}
\end{figure}

Exchange of atoms between the bulb of the cell and the stem with the
Rb reservoir (Fig. \ref{Fig:exp_setup}) can also produce rapid
relaxation and must be mitigated.  This can be accomplished by
employing a ``lockable stem" \cite{Karaulanov2009} which provides a
coated barrier to reduce the rate of exchange between the bulb and
the stem. Finally, gradients of the magnetic field are another
source of relaxation, so care must be taken to minimize them.

In Ref. \cite{Bal2010} it was reported that cells coated with
1-nonadecene ${\rm (CH_2-CH(CH_2)_{16}-CH_2)}$ yielded polarization
lifetimes of about three seconds.  To investigate the alkene-based
coating carefully, we prepared three Rb vapor cells
with lockable stems.  Cells C1 and C2 had natural-abundance Rb and
non-ideal locks, cell C3 had ${\rm ^{87}Rb}$ and a ``precision
ground" lock. The initial material for the coating preparation was
Alpha Olefin Fraction C20-24 from Chevron Phillips (CAS Number
93924-10-8). A light fraction of the material was removed through
vacuum distillation at ${\rm 80^\circ C}$. The remains were used as
the coating material. Coatings were prepared with the procedure
described in Refs. \cite{Ale1992,Bal1995}, except the temperature
was ${\rm 175^\circ C}$ instead of ${\rm 220^\circ C}$. After
preparation, the cells were cured at ${\rm 70^\circ C}$ for several
hours.  Cell C1 had a polarization lifetime of 60 seconds, while
cells C2 and C3 had polarization lifetimes of about 15 seconds.  The
measurements presented here were obtained with cell C1.

The experimental setup is shown in Fig. \ref{Fig:exp_setup}(a).  The
coated cell was placed inside four layers of mu-metal and one layer
of ferrite \cite{Kor2007} shielding.  A circularly polarized pump
beam, propagating in the $z$ direction, tuned near the
$F=2\rightarrow F'$ D1 transitions of $^{85}{\rm Rb}$, optically
pumped the alkali spins. Spin precession was monitored via optical
rotation of linearly polarized probe light, propagating in the $x$
direction, tuned about 1.5 GHz to the blue of the $F=3\rightarrow
F'$ D1 transitions of $^{85}{\rm Rb}$.  Optical rotation, scaling roughly as the inverse of detuning, was dominated by ${\rm ^{85}Rb}$, however there was some contribution from ${\rm ^{87}Rb}$.  Typical probe power was
$\approx {\rm 2~\mu W}$, although much higher probe power could be
used without incurring substantial additional broadening since the
probe was tuned far off resonance.  Pump power ranged from $0-2~{\rm
\mu W}$. Most of the measurements were performed at a temperature of
$30^\circ {\rm C}$ where the Rb vapor density was $ n \approx 1.5
\times 10^{10}~{\rm cm^{-3}}$, measured by transmission of a weak
probe beam. The orientation of the cell could be manipulated from
outside the magnetic shields so that the lock could be opened and
closed without opening the shields. With the lock open, polarization
lifetimes were much shorter than with the lock closed, approximately
3 seconds.  Geometry dictated that the stem and locking bullet were
nearly horizontal, producing a lock of variable quality: the
longitudinal relaxation time varied by as much as a factor of four
from run-to-run.

Using the apparatus shown in Fig. \ref{Fig:exp_setup}(a) we
investigated the relaxation of both the longitudinal and transverse
(with respect to magnetic field) components of spin polarization.
Longitudinal relaxation was measured by first applying a magnetic
field parallel to the pump beam, and then adiabatically rotating the
magnetic field into the direction of the probe beam, and subsequently
monitoring optical rotation of the probe as the longitudinal
polarization decayed. To investigate transverse relaxation, we
observed the transient response of the alkali spins to a
non-adiabatic change in the magnetic field, either by (1) pumping
the spins in zero magnetic field and applying a step in $B_y$, or
(2) by pumping the spins in a finite bias magnetic field $B_z$ and
then applying a short pulse of magnetic field $B_x$, similar to RF
excitation pulses in nuclear magnetic resonance. We also made high
field (10-20 G) measurements of the longitudinal relaxation time
using an apparatus similar to that in Ref. \cite{Gra2005}.

\begin{figure}
  \includegraphics[width=3.4 in]{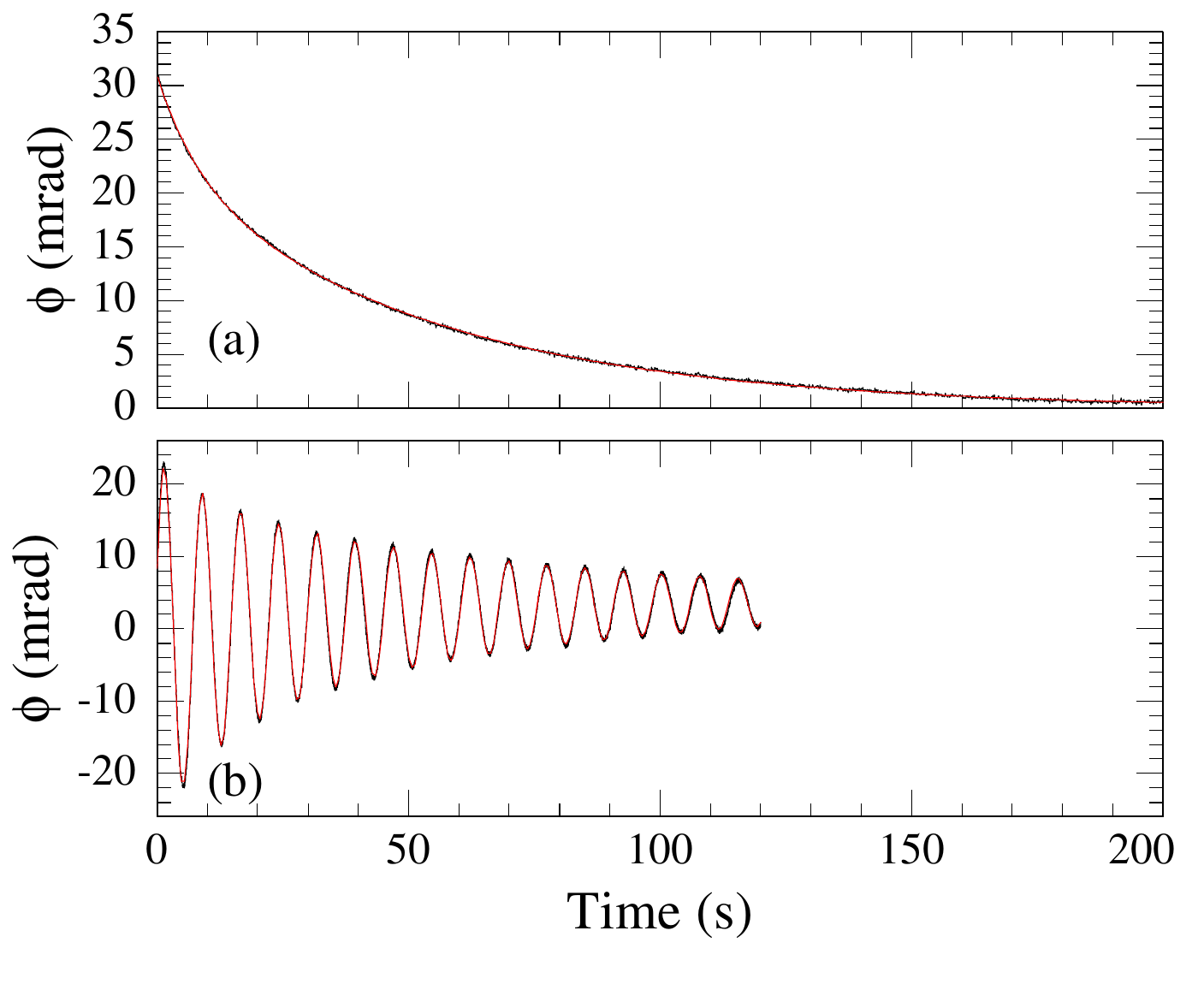}\\
  \caption{(color online) (a) Decay of longitudinal polarization.
  The red trace overlaying the data is a fit to two decaying exponentials, with fast and slow decay times of 8 and 53 seconds, respectively.  (b) Response of alkali spin polarization to a transient
  of the magnetic field.  The fast and slow decay times are 13 and 77 seconds.
  }\label{Fig:Decayplots}
\end{figure}

Figure \ref{Fig:Decayplots} shows optical rotation of the probe due
to longitudinal (a) and transverse (b) polarization when the cell
was performing optimally. The decay of longitudinal polarization is
well described by two exponentials
with fast and slow time constants $T_{1f} = 8~{\rm s}$ and
$T_{1s}=53~{\rm s}$,  respectively.  Such biexponential decays arise
from several competing processes of electron spin-destruction
collisions with the cell walls, residual relaxation due to
collisions with the reservoir, and alkali-alkali spin-exchange
collisions \cite{Gra2005}.

Figure \ref{Fig:Decayplots}(b) shows the transient
response to a step in the magnetic field $B_y\approx 0.2~{\rm \mu
G}$ after pumping at zero magnetic field. In such low magnetic fields, the transient response is well
described (as indicated by the fitted curve overlaying the data) by
an oscillating signal with a single frequency and a return to steady
state, with fast and slow decays characterized by lifetimes $T_{2f}$
and $T_{2s}$.  For the data shown in Fig. \ref{Fig:Decayplots}(b),
$T_{2f} = 13$~s and $T_{2s} = 77$~s.  The presence of only a single
frequency oscillation is because the two isotopes ``lock'' together
in the SERF regime -- in larger magnetic fields we see the
appearance of two frequencies corresponding to free precession of
either isotope in the absence of spin-exchange collisions.   We note
that the low-field measurements shown in Fig. \ref{Fig:Decayplots}
were not repeatable: subsequent values of both $T_{1s}$ and $T_{2s}$
were somewhat shorter, around 25-35 s.  Under these conditions, the
fast relaxation was not apparent, and curves were fit with a single
decay time, $T_{1s}$ or $T_{2s}$.  High field measurements of the
polarization lifetime, with the stem oriented vertically, were
consistently long over the course of several months, in the range of
40-50 seconds. Hence, we suspect that minute-long lifetimes in low
field were difficult to reproduce because of the variable quality of
the locking stem in the nearly horizontal position.

\begin{figure}
  \includegraphics[width=3.4 in]{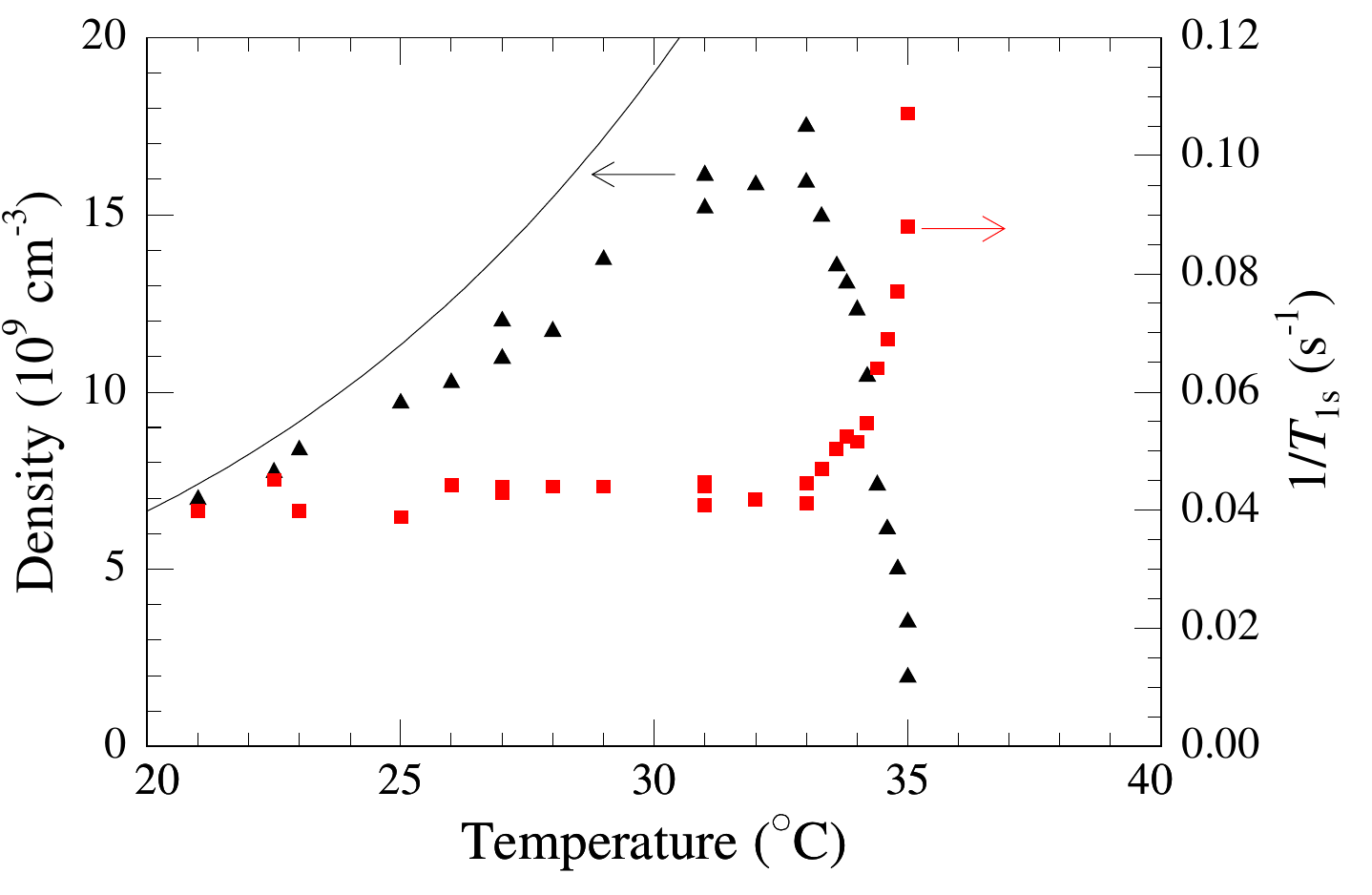}\\
  \caption{(color online) Density (triangles) and relaxation rate (squares) as a function of temperature.
  Density appears to loosely follow that of a saturated vapor (solid curve)
  until it drops rapidly at $33{\rm ^\circ C}$, accompanied by a significant increase in relaxation rate.}\label{Fig:TempDependence}
\end{figure}

The temperature dependences of the density and longitudinal
relaxation rate may be useful in characterizing the properties of
the coating, and are shown in Fig. \ref{Fig:TempDependence}. In
acquiring these data, we let the oven and cell equilibrate for
several hours at each temperature before measuring the density and
lifetime. The lock was closed the entire time.  We see a sharp drop
in density and an increase in relaxation rate as we pass through the
melting transition of the coating at $\approx 33{\rm ^\circ C}$.
This behavior was repeatable and without significant hysteresis, as
data points were acquired non-sequentially. The solid curve represents
the expected density for a saturated vapor \cite{CRC}, which begins
to deviate from the measured density at relatively low temperature.

\begin{figure}
  \includegraphics[width=3.3 in]{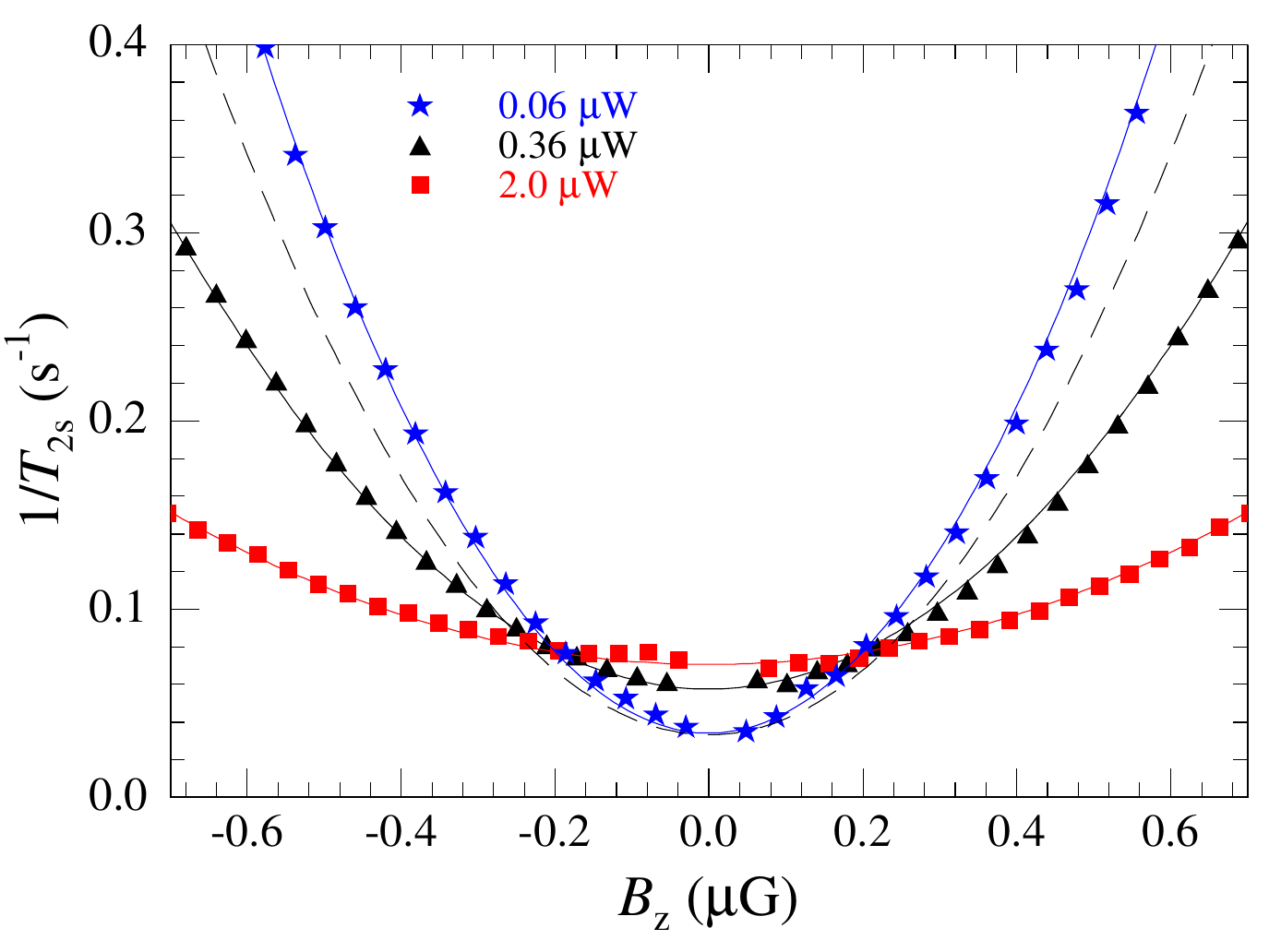}\\
  \caption{(color online) Transverse relaxation rate as a function of magnetic field for several
  values of the pump power. The smooth curves overlaying the data are
  fits to a curve described in the text.  The dashed curve is the
  relaxation rate in the very low polarization limit given by Eq. 1 for nuclear spin $I=5/2$.
  }\label{Fig:Relax_vs_Field}
\end{figure}

We now turn to a discussion of the effects of spin exchange in a
low-density vapor in very low magnetic fields. When the Larmor
precession frequency is small compared to the spin-exchange rate
$1/T_{ex}=n\sigma_{ex}v$ ($n$ is the number density, $\sigma_{ex} =
1.9\times 10^{-14}~{\rm cm^2}$ is the spin-exchange cross section
for Rb \cite{Walter2002}, and $v$ is the mean relative thermal
velocity), spin-exchange collisions produce relaxation that is
quadratic in the magnetic field and modify the effective
gyromagnetic ratio, both of which depend on the degree of spin
polarization \cite{Hap1973}.  Figure \ref{Fig:Relax_vs_Field} shows
the magnetic-field dependence of the transverse relaxation rate,
$1/T_{2s}$, for several pump powers.  For these data, transverse
coherences were produced by applying a short (0.2 s) pulse of
magnetic field in the $x$ direction in the presence of a static
field $B_z$.  The solid curves overlaying the data are fits to
$A+A_{SE}B_z^2$, where $A$ represents relaxation due to wall
collisions, pump light, and gradients, and $A_{SE}$ represents the
contribution to broadening from spin-exchange collisions. Relaxation
deviates from the quadratic behavior shown in Fig.
\ref{Fig:Relax_vs_Field} as the magnetic field is increased,
reaching an asymptotic level of $1/T_{2s}\approx 3~{\rm s^{-1}}$ at
$\approx$ 1 mG.  At low magnetic field, increasing pump power
produces power broadening, however, at higher magnetic fields, high
pump power reduces spin-exchange relaxation by preferentially
populating the stretched state, which is immune to spin-exchange
relaxation \cite{App1999}.

For a single isotope with a spin-temperature distribution and low
polarization, spin-exchange relaxation is given by \cite{Hap1977}
\begin{equation}\label{Eq:lowpolrelaxation}
    \Gamma_{SE} = \omega_0^2T_{ex}\frac{q_0^2-(2I+1)^2}{2q_0},
\end{equation}
where $\omega_0 = g_s\mu_B B/q_0\hbar$, $I$ is the nuclear spin, and
$q_0 = [I(I+1)+S(S+1)]/S(S+1)$ is the nuclear slowing-down factor. The dashed
curve in Fig. \ref{Fig:Relax_vs_Field} is the expected relaxation
rate in the low polarization limit for a vapor of pure ${\rm
^{85}Rb}$ $(I=5/2)$ given by Eq. \eqref{Eq:lowpolrelaxation}, with $T_{ex}$
determined by transmission measurements of the density. Equation
\eqref{Eq:lowpolrelaxation} appears close to accounting for
relaxation at low light power, however there is clearly some
discrepancy, presumably due to the presence of two isotopes.

\begin{figure}
  \includegraphics[width=3.4 in]{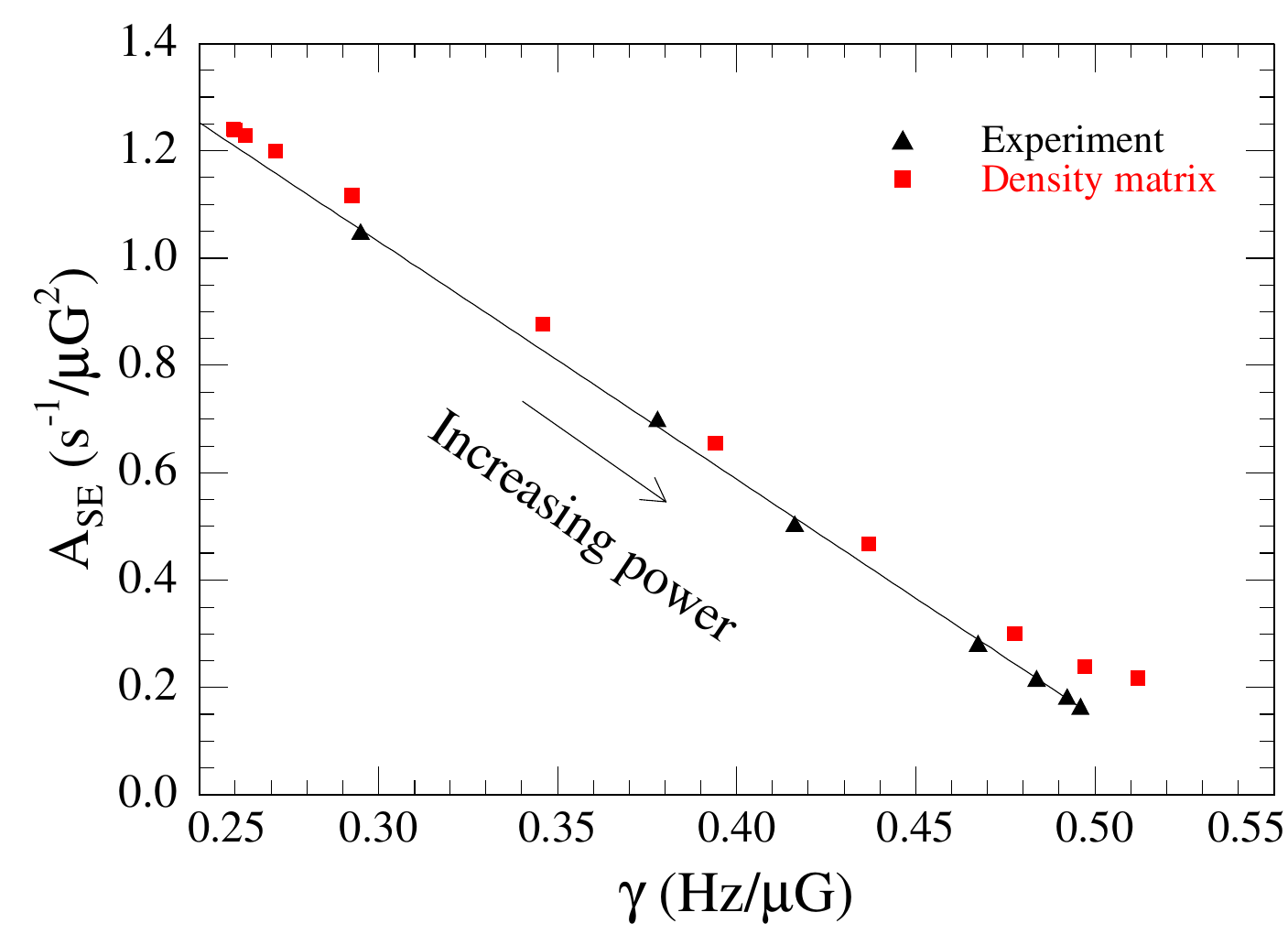}\\
  \caption{(color online) Triangles show experimental measurements of spin-exchange broadening $A_{SE}$ vs. the effective gyromagnetic ratio $\gamma$ for pump power ranging from $0.06~{\rm \mu W}$ to ${\rm 2~\mu W}$.  The straight line overlaying the data is a linear fit.  Squares show the results of density matrix calculations.}\label{Fig:gamma_broad_theory_exp}
\end{figure}

The gyromagnetic ratio also varies significantly with pump power. In order to compare with theoretical calculations (see below) it is convenient to plot the measured spin-exchange broadening $A_{SE}$ as a function of the effective gyromagnetic ratio $\gamma$ (Fig. \ref{Fig:gamma_broad_theory_exp}, triangles).  It is interesting to note that there is a linear relationship between these two parameters, as indicated  by the linear fit overlaying the data. It is also worth noting that, in these measurements, spin-exchange broadening approaches an asymptotic value of about $0.2~{\rm s^{-1}/\mu G^2}$ at high power due to the presence of two isotopes, as can be seen by the clustering of data points at high light power, despite the increasing size of light power steps. In an isotopically pure vapor, relaxation due to spin-exchange collisions could be largely eliminated at high pump power by hyperfine pumping, similar to the light narrowing observed in Ref. \cite{App1999}.

To further investigate the effects of spin-exchange collisions, we
performed numerical simulations following the approach used in
Refs. \cite{Appelt1997,Savukov2005}.  The contributions to evolution
of the ground state density matrix $\rho_j$ for isotope $j$ due to
hyperfine splitting, Zeeman splitting, optical pumping, spin-destruction,
and spin-exchange are, respectively:
\begin{eqnarray}\label{Eq:DM1}
 \nonumber   \frac{d\rho_j}{dt} &=& \frac{a_{j}}{i\hbar}[\mathbf{I}_j\cdot\mathbf{S}_j,\rho_j] + \frac{g_s\mu_B}{i\hbar}[\mathbf{B}\cdot\mathbf{S}_j,\rho_j]\\
 \nonumber &+&R\lbrack\phi_{85}(1+2\mathbf{\hat{z}}\cdot\mathbf{S}_{85})-\rho_j\rbrack+\frac{\phi_j-\rho_j}{T_{sd}} \\
 &+& \sum_k\frac{\phi_j(1+4\langle \mathbf{S}_k\rangle\cdot\mathbf{S}_j)-\rho_j}{T_{ex,jk}}\label{Eq:DM}.
\end{eqnarray}
Here $a_{j}$ is the hyperfine constant, $\mathbf{I}_j$ is the
nuclear spin, $g_s$ is the Land\'e factor for the electron, $\mu_B$
is the Bohr magneton, $R$ is the optical pumping rate for ${\rm
^{85}Rb}$ (there is no optical pumping of ${\rm ^{87}Rb}$ since the
pump light is resonant only with ${\rm ^{85}Rb}$ transitions), and
$\phi_j=\rho_j/4+\mathbf{S}_j\cdot\rho_j\mathbf{S}_j$ is the purely
nuclear part of the density matrix. The spin-destruction rate
$T_{sd}$ is determined from measurements of $T_1$, and the
spin-exchange rates $T_{ex,jk}$ are determined by the measured
alkali density and the known cross-sections.  The transient response
to a pulse of magnetic field in the $y$ direction and subsequent
precession around a static field in the $z$ direction is determined
by numerically integrating Eq. \eqref{Eq:DM1}, starting from a
spin-temperature distribution along the $z$ axis.  We extract the
$x$ component of electron spin polarization, weighted by isotopic
abundance $\eta_j$, $\langle S_x\rangle = \eta_{85} \langle
S_{x,85}\rangle+\eta_{87} \langle S_{x,87}\rangle$, a reasonable
approximation of the experimental observable, and fit this to a
decaying sinusoid.  The squares in Fig.
\ref{Fig:gamma_broad_theory_exp} show the results of simulations.
Experiment and simulation are in good agreement for low light power,
although there is some small systematic offset, which we attribute
to uncertainty in the alkali vapor density.  At higher light power,
the simulation deviates from experiment, presumably because the
optical pumping term in Eq. \eqref{Eq:DM1} is correct in the limit
of unresolved hyperfine structure, and therefore cannot account for
hyperfine pumping present in the experiment.

In conclusion, we have demonstrated that an alkene coating can
support up to $10^6$ alkali-wall collisions before depolarizing the
alkali spins when all other sources of relaxation are properly
mitigated.  This represents an improvement by nearly a factor of 100
over traditional coatings. We demonstrate here that cells employing
such a coating can enable operation of a SERF magnetometer in a room
temperature environment, dramatically expanding the scope of
applications for such magnetometers.  In addition to magnetometry,
anti-relaxation coatings are used in a number of other contexts in
both pure and applied research.  As we outlined here, alkene
coatings can be used to study the effects of spin-exchange
collisions in very low density environments, and may be of use for
investigating more subtle atom-atom collisions.  Alkali vapor cells
utilizing such coatings may also dramatically improve the
performance of atomic clocks, depending on the nature of the
hyperfine shifts associated with atom-wall collisions, a subject of
future investigation.  Alkene coated cells may greatly enhance the
lifetime of quantum memory applications \cite{Jul2004} or the
storage time of light in ``slow-light" experiments \cite{Bud1999,
Kle2006}. In the context of geophysical measurements, extremely
narrow lines can reduce orientation dependent ``heading errors'' due
to the non-linear Zeeman effect, a significant issue in geomagnetic
surveying \cite{Ale2003}. While spin-exchange relaxation is
difficult to completely eliminate at high field, the use of only a
single isotope and hyperfine pumping may reduce such relaxation
considerably. This work was supported by the ONR MURI program and by
the NSF.

\end{document}